\documentclass{ecai}  
\pdfoutput=1
\usepackage{graphicx}
\usepackage{latexsym}
\usepackage{bchart}
\usepackage{pgfplots}
\pgfplotsset{compat=1.16}
\usepackage{caption}
\usepackage{url}
\ecaisubmission      

\usepackage{hyperref}
\hypersetup{
  colorlinks   = true,    
  urlcolor     = blue,    
  linkcolor    = blue,    
  citecolor    = red      
}

\usepgfplotslibrary{colorbrewer} 

\colorlet{mgrefColor-1}{Paired-B}
\colorlet{mgrefColor-2}{Paired-F}
\colorlet{mgrefColor-3}{Paired-D}
\colorlet{mgrefColor-4}{Paired-H}
\colorlet{mgrefColor-5}{Paired-J}
\colorlet{mgrefColor-6}{Paired-L}
\colorlet{mgrefColor-7}{Set2-D}
\colorlet{mgrefColor-8}{Dark2-H}
\colorlet{mgrefColor-9}{Set2-E}

\colorlet{mgrefColorLight-1}{Paired-A}
\colorlet{mgrefColorLight-2}{Paired-E}
\colorlet{mgrefColorLight-3}{Paired-C}
\colorlet{mgrefColorLight-4}{Paired-G}
\colorlet{mgrefColorLight-5}{Paired-I}
\colorlet{mgrefColorLight-6}{Paired-K}
\colorlet{mgrefColorLight-7}{Dark2-D}
\colorlet{mgrefColorLight-8}{Set2-H}
\colorlet{mgrefColorLight-9}{Dark2-E}

\pgfplotscreateplotcyclelist{mgrefColorLight}{%
	mgrefColorLight-1,mgrefColorLight-2,mgrefColorLight-3,mgrefColorLight-4,mgrefColorLight-5,mgrefColorLight-6,mgrefColorLight-7,mgrefColorLight-8,mgrefColorLight-9}	 

\pgfplotscreateplotcyclelist{mgrefColor}{%
	mgrefColor-1,mgrefColor-2,mgrefColor-3,mgrefColor-4,mgrefColor-5,mgrefColor-6,mgrefColor-7,mgrefColor-8,mgrefColor-9}

\pgfplotscreateplotcyclelist{mgrefColorFilled}{%
	{mgrefColor-1,fill=mgrefColorLight-1},
	{mgrefColor-2,fill=mgrefColorLight-2},
	{mgrefColor-3,fill=mgrefColorLight-3},
	{mgrefColor-4,fill=mgrefColorLight-4},
	{mgrefColor-5,fill=mgrefColorLight-5},
	{mgrefColor-6,fill=mgrefColorLight-6},
	{mgrefColor-7,fill=mgrefColorLight-7},
	{mgrefColor-8,fill=mgrefColorLight-8},
	{mgrefColor-9,fill=mgrefColorLight-9}
}

\begin{document}

\begin{frontmatter}

\title{Performance of Large Language Models \\
in a Computer Science Degree Program}

\author{\fnms{}~\snm{Tim Krüger}\thanks{tim.krueger@stud.hn.de}}
\author{\fnms{}~\snm{Michael Gref}\thanks{michael.gref@hs-niederrhein.de}} 

\address{Niederrhein University of Applied Sciences, Krefeld}


\begin{abstract}
Large language models such as ChatGPT-3.5 and GPT-4.0 are ubiquitous and dominate the current discourse. Their transformative capabilities have led to a paradigm shift in how we interact with and utilize (text-based) information. Each day, new possibilities to leverage the capabilities of these models emerge.

This paper presents findings on the performance of different large language models in a university of applied sciences' undergraduate computer science degree program. Our primary objective is to assess the effectiveness of these models within the curriculum by employing them as educational aids. By prompting the models with lecture material, exercise tasks, and past exams, we aim to evaluate their proficiency across different computer science domains.

We showcase the strong performance of current large language models while highlighting limitations and constraints within the context of such a degree program. We found that ChatGPT-3.5 averaged 79.9\% of the total score in 10 tested modules, BingAI achieved 68.4\%, and LLaMa, in the 65 billion parameter variant, 20\%. Despite these convincing results, even GPT-4.0 would not pass the degree program - due to limitations in mathematical calculations.
\end{abstract}

\end{frontmatter}

\section{Introduction}
In the realm of natural language processing, large language models, hereafter only referenced as LLMs, have now become an integral part of our digital landscape. They have a widespread influence in today's discourse and a ubiquitous presence in various fields and industries \cite{AI_statista}. Among these models, ChatGPT-3.5 and GPT-4.0 have emerged as prominent examples, captivating the attention of students, researchers, and developers alike.

It is essential to look at these models in the context of higher education because they provide new ways and possibilities to teach, learn and perceive information. Useful for both students and instructors. They could help students, for example, by delivering a more personalized and interactive educational experience and acting as a kind of "learning buddy." For an instructor, the possibilities are also plenty. These models can generate supplementary materials, explanations, or examples \cite{Gimpel2023Unlocking}. Alternatively, they could aid in the assessment process by automating the grading procedure for all text-based requirements.

A lot of research is currently taking place on this topic. For example, H. Gimpel et al. \cite{Gimpel2023Unlocking} have written an extensive essay on the opportunities but also the risks that generative AI models bring to higher education by collecting nearly 50 high-quality scholarly sources. They provide guidance for both students and instructors by providing hands-on recommendations for the usage of AI in higher education.

ChatGPT-3.5 and GPT-4.0 are not the only AI models impacting learning and teaching; much more software exists. \href{https://www.deepl.com/write}{DeepL Write} can improve writing, from fixing grammar and punctuation mistakes to rephrasing entire sentences or sections. The same is true for \href{https://app.grammarly.com/}{Grammarly}, which offers users an AI text generation functionality for further improvements and suggestions regarding clarity, engagement, and delivery of a text. Even the creation of multimedia content is no problem. Programs like \href{https://www.midjourney.com/home}{Midjourney} and \href{https://openai.com/dall-e-2}{Dall-E} allow users the creation of photorealistic images and visualizations with just a few prompts \cite{kruger2022kunstliche}. Furthermore, when Microsoft releases \href{https://blogs.microsoft.com/blog/2023/03/16/introducing-microsoft-365-copilot-your-copilot-for-work/}{Copilot}, their AI support tool, with Office \cite{MS_copilot}, the use of AI will have also arrived in all non-technical disciplines. These tools will then be used passively daily by millions of people, so we must look at the opportunities but also threats that these technologies can bring to higher education and learning/teaching in general.

As part of the research for this paper, we interviewed several professors from our teaching institution. We identified cheating and plagiarism as one of the main concerns. H. Gimpel et al. \cite{Gimpel2023Unlocking} have also dedicated several pages of their essay to this topic and stressed the importance of rules and guidelines that should be in place for the university environment without denying students access to this new technology.
However, as the present work is limited to evaluating the performance of LLMs, this is an aspect to be explored in subsequent work.

To accurately assess the benefits of this technology and their usage as educational aids, we set out to evaluate the performance across our undergraduate computer science curriculum.
In total, we collected 40 data points, where one data point represents the performance of one LLM or LLM variant in one module of the degree program.

\section{Related Work}
In this section, we want to explore some of the various research efforts that have examined the performance of LLMs in the field of computer science. The results, some of which differ significantly, inspired us to test the performance of these LLMs in our degree program as well.
Table \ref{table:Performance_CS} shows a small selection of test and exam results published by OpenAI \cite{openai2023gpt4}, with the release of their GPT-4.0 model, and one exam (Algorithms and Data Structures) tested by Bordt et al. \cite{bordt2023chatgpt}.

The first results are from LeetCode, a popular online platform that provides programming exercises and coding challenges commonly found in technical interviews \cite{harper2022interview}.
The platform is aimed at software developers and programmers to enhance their programming skills by solving algorithmic problems \cite{enwiki:1159537770}.
The programming exercises and algorithmic problems are divided into three difficulty ranges (easy, medium, and hard) \cite{nikolaidisend}.

\begin{table}[t]
	\begin{center}
		{\caption{Exam results for GPT-3.5 and GPT-4.0 when\\
				tested by OpenAI \cite{openai2023gpt4} and Bordt et al. \cite{bordt2023chatgpt}.\\
				Values are rounded to the first decimal place.}\label{table:Performance_CS}}
		\begin{tabular}{ccc}
			\hline
			\\[-6pt]
			\textbf{Test / Exam} & \textbf{GPT-3.5} & \textbf{GPT-4.0} \\
			\hline
			\\[-6pt]
			\quad LeetCode (Hard) 				& 0.0\% 	& 6.6\% \\
			\quad LeetCode (Easy) 				& 29.3\% 	& 75.6\% \\
			\quad Algorithms \& Data Structures & 51.3\% 	& 60.0\% \\
			\hline
		\end{tabular}
	\end{center}
\end{table}

In the \textit{easy} problem section, GPT-3.5 answered 12 out of 41 questions correctly, resulting in a performance of 29.3\%. GPT-4.0 answered 31 out of 41 questions correctly, resulting in a performance of 75.6\% - an improvement of 47.3 percentage points \cite{openai2023gpt4}.
The results may depend on the exact category and programming language \cite{nikolaidisend}. Nikolaidis et al. found that in their case, ChatGPT-3.5 solved 45\% of 50 randomly selected \textit{easy} LeetCode problems correctly while providing noticeably better results in the programming languages Java and Python.

When tested by OpenAI, GPT-3.5 could not solve a single of the \textit{hard} problems on LeetCode \cite{openai2023gpt4}. These results again may depend on the type of problem that had to be solved \cite{nikolaidisend}. Nikolaidis et al. found that ChatGPT-3.5 solved 10 out of 21 \textit{hard} problems correctly, resulting in 47.6\% accuracy. ChatGPT-3.5 would then, in fact, even severely outperform GPT-4.0 when tested by OpenAI, which was able to solve 3 out of 45 \textit{hard} problems correctly (6.6\% accuracy) \cite{openai2023gpt4}.

Bordt et al. tested ChatGPT-3.5 and GPT-4.0 on an undergraduate computer science exam in \textit{Algorithms and Data Structures}. The exam was fed to the LLMs in the same way students would receive it. The answers of the models were transferred to paper by the testers and mixed with the solutions of the students \cite{bordt2023chatgpt}. ChatGPT-3.5 scored 20.5 out of 40 possible points (51.25\%), allowing it to pass the exam narrowly. GPT-4.0 improved that score by 8.75 percentage points, reaching 60\% (24/40 points). With this result, GPT-4.0 outperforms the average student, which scores 23.9 in the mean \cite{bordt2023chatgpt}.

Both ChatGPT-3.5 and GPT-4.0 indicate wide-ranging capabilities in the field of computer science. GPT-4.0 also seems to be an improvement over GPT-3.5 in every way. The findings on performance variation are worth noting for our research. The LLMs' answers seem to depend on the corresponding computer science discipline category and the specific programming language asked \cite{nikolaidisend}. It is also relevant to note that the programming errors generated seem to be mainly semantic. The models hardly make syntax errors, but the code, if wrong, can have serious logic errors \cite{nikolaidisend}.

\section{Methodology}
The crux of our methodology is the evaluation of various LLMs by feeding them academic content drawn from a bachelor's degree program in computer science at a university of applied sciences. We aim to determine each model's overall performance and identify the highest and lowest-scoring modules, grade distributions, and potential affinities for certain topics. Additionally, the study aims to determine whether the models would complete the degree program.

Our data set comprised samples of past exams from ten different modules of the degree program, see Table \ref{table:Test_data_set}. This core data set was complemented with information from questionnaires, practice exercises, and lecture notes to offer a more holistic view of the curriculum. For modules with oral exams, the questions were based on the same data but created in consultation with the supervising professor to simulate realistic exam scenarios. Only examinations for which the professors gave their approval were taken into account for the study.

The criteria for evaluation were adapted for each module. In written exams, we employed the evaluation system and point allocation provided by the supervising professor. In oral exams, we weighted questions according to complexity and difficulty. These questions were finalized in consultation with the supervising professors. Evaluating and assessing the performance involved verifying correctness, compiling and testing program code, and recalculating solutions.

Due to the limitations of certain models in handling multimedia input, we partly excluded those tasks from our assessment.
Adjustments were made to the total score and weighting of the exam accordingly. In instances where it was feasible, we transformed such tasks into a suitable textual format with tables and data structures being converted to markdown and diagrams re-imagined into the corresponding UML representation.

\begin{table}[t]
	\begin{center}
		{\caption{Overview of our test data set.}\label{table:Test_data_set}}
		\begin{tabular}{ccc}
			\hline
			\\[-6pt]
			\textbf{Semester} & \textbf{Written exams} & \textbf{Oral exams} \\
			\hline
			\\[-6pt]
			\quad1 & 0 & 0 \\
			\quad2 & 2 & 0 \\
			\quad3 & 2 & 0 \\
			\quad4 & 2 & 1 \\
			\quad5 & 2 & 1 \\
			\hline
			\\[-6pt]
			\quad{Sum} & 8 & 2\\
			\hline
		\end{tabular}
		\vspace{-12pt}
	\end{center}
\end{table}

The prompting of the models was a carefully considered aspect of this research project. Prompt engineering has been shown to improve the performance of models in various studies (e.g. \cite{white2023prompt}). However, to provide a broad overview of the performance across the curriculum, we opted to prompt all models only once and use the first response provided by each model. Before starting the assessment, a generic pre-prompt was given in each case, setting the context that they were interacting in a simulated exam scenario and outlining expectations for responses.\\

\begin{minipage}{\linewidth}
\begin{quote}
\textit{\hspace{1em}I am now going to ask you a few questions from a hypothetical [insert topic or subject] exam of an undergraduate computer science degree program. I want you to answer the questions to the best of your knowledge and capabilities. Please answer briefly and concisely unless I explicitly ask for a more detailed answer! Please answer purely in continuous text or bullet points. If output in chart or table form is desired, I will let you know.}
\end{quote}
\end{minipage}
\vspace{0.2cm}

We tested ChatGPT-3.5, BingAI, which uses the GPT-4.0 foundation model, StableLM-Alpha in the 7 billion parameter version, and LLaMa in both the 7 billion and 65 billion parameter versions. Towards the end of our project, we also received access to GPT-4.0 but were restricted in using this model due to time constraints. We viewed these selections as an appropriate mix of open- and closed-access LLMs.

StableLM includes various LLMs published by Stability AI. The size of these models ranges from 3 billion to 65 billion parameters. A 175 billion parameter variant is also planned \cite{stableLM_github}. The models are published in different versions and trained on different datasets. We use the StableLM-Alpha-7B variant, which was trained on a dataset based on \textit{The Pile} \cite{pile}. All models are hosted on \textit{The Hugging Face Hub}, and some are accessible through a web interface \cite{stableLM_hugging_face}.

LLaMa refers to a collection of different LLMs ranging from 7B to 65B parameters, published by MetaAI \cite{touvron2023llama}.
We used LLaMa with the project \textit{llama.cpp}, an open-source C/C++ port of several LLMs \cite{llama.cpp}. This project supports 8-bit, 5-bit, and 4-bit integer quantization, a technique that significantly reduces the memory requirements of the models. In the case of LLaMa, this allowed us to run the models in RAM instead of GPU memory. We considered this approach a more realistic simulation, as the models otherwise require a significant amount of GPU memory. However, there is the possibility of a degradation in model accuracy. There seems to be a trade-off between model size and quality, depending on the quantization method \cite{yao2023zeroquantv2}. In the case of LLaMa-7B, the file size got reduced from 13 GB when using 16-bit floats to 3.5 GB when using 4-bit integer quantization. The perplexity \cite{gonen2022demystifying} rose from 5.9066 to 6.1565, an increase of only 4.23\% \cite{llama.cpp}.

\section{Experimental Results}
We tested ten modules each with ChatGPT-3.5 and BingAI, six modules with GPT-4.0, and fourteen modules in total with StableLM-Alpha-7B, LLaMa-7B, and LLaMa-65B, resulting in forty data points. We have not been able to test every model iteration on every module of the curriculum due to the time constraints of this project. However, the following data underpins what we present as a comprehensive insight into the performance of these models across an array of computer science curriculum modules.

Referring to grades in the following, we calculated them according to the \textit{modified Bavarian formula} corresponding to the German grading system \cite{modified_bavarian_formula}. Depending on the university, a conversion may be necessary. If not stated otherwise, 50\% of the score are required to pass the exam.

\subsection{1st Semester}
We have neither received full approval nor the required content for any of the modules of the first semester from the responsible professors. This is left to be explored in subsequent work.

\subsection{2nd Semester}
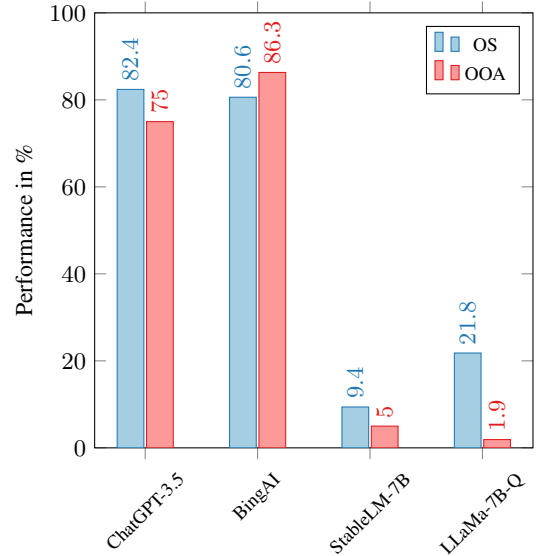
\begin{figure}[t]
\centering
\begin{tikzpicture}
\begin{axis}[
    ybar=2*\pgflinewidth,
    symbolic x coords={ChatGPT-3.5, BingAI, StableLM-7B, LLaMa-7B-Q},
    enlarge x limits=0.15,
    ybar=1pt,
    xtick=data,
    ylabel=Performance in \%,
    nodes near coords,
    nodes near coords style={rotate=90, anchor=west},
    ymin=0,
    ymax=100,
    bar width=10pt,
    cycle list name = mgrefColorFilled,
    width=0.85\linewidth,
    height=0.85\linewidth,
    legend style={font=\scriptsize},
    legend pos = north east,
    xticklabel style={font=\scriptsize, rotate=45, align=center},
    xlabel style={align=center}
]
\addplot coordinates {(ChatGPT-3.5, 82.4) (BingAI, 80.6) (StableLM-7B, 9.4) (LLaMa-7B-Q, 21.8)};
\addplot coordinates {(ChatGPT-3.5, 75.0) (BingAI, 86.3) (StableLM-7B, 5.0) (LLaMa-7B-Q, 1.9)};
\legend{OS, OOA}
\end{axis}
\end{tikzpicture}
\caption{Exam results for Operating Systems (OS) and \\
Object Oriented Application Development (OOA)}
\label{fig:Semester_2_Performance}
\end{figure}

We have received approval for two second-semester modules, Operating Systems, and Object-Oriented Application Development. Figure \ref{fig:Semester_2_Performance} shows the exam results for each LLM in these modules.\\

Operating Systems (OS) is a five credit-point module. In the module, students learn the structure of a modern operating system and algorithms and strategies for managing and allocating resources in it. They also develop programs in a UNIX environment and work out solutions to problems of interprocess communication \cite{modulhandbuch_po2013}.

Object Oriented Application Development (OOA) is a seven credit-point module. It focuses on teaching the methods and techniques of object-oriented programming. Requirements are implemented using efficient algorithms and data structures. Programming is done in C++  \cite{modulhandbuch_po2013}.\\

ChatGPT-3.5 and BingAI performed quite well in OS, scoring 82.4\% and 80.6\%, respectively. The 7B parameter models performed significantly worse. LLaMa-7B-Q (\textbf{q}uantized) scored 21.8\%, and StableLM-7B scored 9.4\%. While StableLM-7B could answer almost no questions, LLaMa-7B-Q could still answer questions about shell commands and general operating system terms. Nevertheless, it was too little to pass the exam. ChatGPT-3.5 and BingAI were able to answer many questions. The models did make mistakes when calculating memory usage and applying paging algorithms. This cost them points but kept the good result the same. ChatGPT-3.5 passed this exam with a grade of 2.0 and BingAI with a Grade of 2.1.

OOA is the only module in which BingAI performed better than ChatGPT-3.5 in all our testing. The latter LLM scored 75\%, whereas BingAI scored 86.3\%. This results in grades of 2.5 for ChatGPT-3.5 and 1.8 for BingAI. The score for BingAI is one of the best results for this LLM in all our tests. The biggest problems ChatGPT-3.5 had were with implementing the object-oriented interfaces in C++. The code compiled but either didn't work as it should or implemented something completely different from the task. StableLM-7B performed slightly better than LLaMa-7B-Q in this exam. However, both LLMs had severe problems with the assignments, solving almost no tasks.

\subsection{3rd Semester}
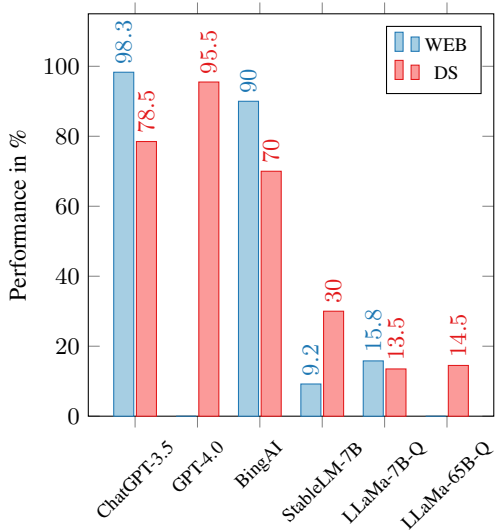
\begin{figure}[t]
\centering
\begin{tikzpicture}
\begin{axis}[
    ybar=2*\pgflinewidth,
    symbolic x coords={ChatGPT-3.5, GPT-4.0, BingAI, StableLM-7B, LLaMa-7B-Q, LLaMa-65B-Q},
    enlarge x limits=0.15,
    ybar=1pt,
    xtick=data,
    ylabel=Performance in \%,
    nodes near coords,
    nodes near coords style={rotate=90, anchor=west},
    ymin=0,
    ymax=115,
    bar width=7.5pt,
    cycle list name = mgrefColorFilled,
    width=0.80\linewidth,
    height=0.80\linewidth,
    legend style={font=\scriptsize},
    legend pos = north east,
    xticklabel style={font=\scriptsize, rotate=45, align=center},
    xlabel style={align=center},
    nodes near coords={%
       \pgfmathfloatifflags{\pgfplotspointmeta}{0}{}{\pgfmathprintnumber{\pgfplotspointmeta}}},
]
\addplot coordinates {(ChatGPT-3.5, 98.3) (GPT-4.0, 0) (BingAI, 90) (StableLM-7B, 9.2) (LLaMa-7B-Q, 15.8) (LLaMa-65B-Q, 0)};
\addplot coordinates {(ChatGPT-3.5, 78.5) (GPT-4.0, 95.5) (BingAI, 70) (StableLM-7B, 30.0) (LLaMa-7B-Q, 13.5) (LLaMa-65B-Q, 14.5)};
\legend{WEB, DS}
\end{axis}
\end{tikzpicture}
\caption{Exam results for Web-Engineering (WEB)\\
and Distributed Systems (DS)}
\label{fig:Semester_3_Performance}
\end{figure}

We have received approval for two modules of the third semester, Web Engineering and Distributed Systems.
The results for the modules of the third semester can be seen in Figure \ref{fig:Semester_3_Performance}.\\

Web Engineering (WEB) is a five credit-point module. It covers the technical fundamentals of modern web-based technologies and architectural, development, and analysis tools for web-based systems. On the front end, students in this module work with HTML, CSS, and JavaScript; On the backend side, with a mixture of Javascript and Python \cite{modulhandbuch_po2013}. A student with 33\% or more would pass the exam, as determined by the supervising professor.

Distributed Systems (DS) is also a five credit-point module. Students of this module learn about distributed system architectures and techniques for synchronization and communication. At the end of this module, they can design, implement and evaluate their own distributed computing structures. The implementation within this module takes place in C/C++ \cite{modulhandbuch_po2013}. Our set exam consists of questionnaire material.\\

WEB was one of the best exam results in all our testing for ChatGPT-3.5, scoring an even 1.0 on the exam with 98.3\%. Even the most extensive task, a partial Python implementation of a backend server for the membership management of a business, was solved completely and correctly. BingAI was also able to answer almost every question correctly. Only in the implementation part did BingAI make logical errors and omit required functionalities. This still resulted in 90\% or a grade of 1.4. StableLM-7B and LLaMa-7B-Q had surprisingly massive problems in this exam, despite the extensive question part. Almost no question could be answered completely or correctly. Both models also failed the implementation part. LLaMa-7B-Q scored 15.8\%, slightly better than StableLM-7B, with 9.2\%.

In DS, both ChatGPT-3.5 and BingAI performed worse than in WEB. ChatGPT-3.5 got a grade of 2.2 with a result of 78.5\%, and BingAI a 2.8 with 70\%. The models could answer almost all simple or introductory questions to the topic correctly but had problems with more in-depth questions, e.g., on network data formats or broker implementations. DS was the first module in which we tested GPT-4.0. With a result of 95.5\%, it got a grade of 1.2 and topped the grade of ChatGPT-3.5 by a whole level. GPT-4.0 answered almost every question completely and correctly in this exam. StableLM-7B scored 30\% in DS, the best result for this LLM in all our tests. Surprisingly, it was able to answer difficult questions on \textit{CORBA}, \textit{SOAP} interfaces, and synchronization mechanisms but failed on simpler, more general questions, like resilience and fault tolerance of distributed systems. Otherwise, it would have had a real chance to pass the exam. In this module, we also tested LLaMa-65B-Q for the first time. With a result of 14.5\%, it performed only slightly better than LLaMa-7B-Q with 13.5\%. Considering the difference in size, this is a disappointing result.

\subsection{4th Semester}
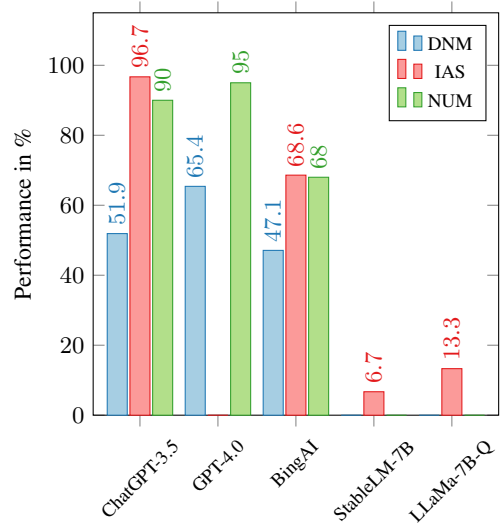
\begin{figure}[t]
\centering
\begin{tikzpicture}
\begin{axis}[
    ybar=2*\pgflinewidth,
    symbolic x coords={ChatGPT-3.5, GPT-4.0, BingAI, StableLM-7B, LLaMa-7B-Q},
    enlarge x limits=0.15,
    ybar=1pt,
    xtick=data,
    ylabel=Performance in \%,
    nodes near coords,
    nodes near coords style={rotate=90, anchor=west},
    ymin=0,
    ymax=115,
    bar width=7.5pt,
    cycle list name = mgrefColorFilled,
    width=0.80\linewidth,
    height=0.80\linewidth,
    legend style={font=\scriptsize},
    legend pos = north east,
    xticklabel style={font=\scriptsize, rotate=45, align=center},
    xlabel style={align=center},
    nodes near coords={%
       \pgfmathfloatifflags{\pgfplotspointmeta}{0}{}{\pgfmathprintnumber{\pgfplotspointmeta}}},
]
\addplot coordinates {(ChatGPT-3.5, 51.9) (GPT-4.0, 65.4) (BingAI, 47.1) (StableLM-7B, 0) (LLaMa-7B-Q, 0)};
\addplot coordinates {(ChatGPT-3.5, 96.7) (GPT-4.0, 0) (BingAI, 68.6) (StableLM-7B, 6.7) (LLaMa-7B-Q, 13.3)};
\addplot coordinates {(ChatGPT-3.5, 90) (GPT-4.0, 95) (BingAI, 68) (StableLM-7B, 0) (LLaMa-7B-Q, 0)};
\legend{DNM, IAS, NUM}
\end{axis}
\end{tikzpicture}
\caption{Exam results for Data Network Management (DNM), Interactive Systems (IAS) and Numerical Analysis (NUM)}
\label{fig:Semester_4_Performance}
\end{figure}

We have received approval for three modules of the fourth semester: Data Network Management, Interactive Systems, and Numerical Analysis. The results can be taken from Figure \ref{fig:Semester_4_Performance}
\pagebreak

Data Network Management (DNM) is a six credit-point module. It provides in-depth, application-oriented knowledge of network administration. Students in this module acquire skills in the design, development, and deployment of large-scale computer networks, as well as techniques for securing them \cite{modulhandbuch_po2013}.

Interactive Systems (IAS) is a five credit-point module. It focuses on software ergonomics and the design and implementation of portable interactive systems. Students of this module learn how to model application-oriented and ergonomic human-machine interfaces \cite{modulhandbuch_po2013}. The implementations in this module are web-based in the programming languages JavaScript and Python. As determined by the supervising professor, the module is considered to be passed if 33\% of the total points are achieved.

Numeric Analysis (NUM) is an elective course in our computer science bachelor's degree program, which gives five credit points. Topics covered include computer arithmetic and rounding errors, systems of linear equations, and linear equilibrium calculus. The module is concluded with an oral examination \cite{modulhandbuch_po2013}. This is one of the modules in which we simulated an exam by taking questions from a questionnaire.\\

DNM is the first module in our tests in which even the larger LLMs have experienced problems. ChatGPT-3.5 barely passed the exam with 51.9\% or a grade of 3.8. BingAI had even more difficulties and failed the exam with a score of only 47.1\%. The application of firewall rules and routing protocols for custom multi-area networks presented in the exam was particularly problematic for both LLMs.
GPT-4.0 performed the best in this exam. It was also unable to completely solve the more difficult tasks but often provided correct partial solutions or made less serious errors than the other two LLMs. GPT-4.0 passed the exam with 65.4\% or a grade of 3.0.

IAS went very well for ChatGPT-3.5. With 96.7\%, it got a grade of 1.1. It answered almost all comprehension and knowledge questions correctly. Even more complex tasks, such as the design of a user interface, were solved completely and correctly. BingAI performed almost 30 percentage points worse in this exam, resulting in one of the biggest gaps between these two LLMs in all our testing. It got a grade of 2.4, or 68.8\% of the total score. The grade of 2.4 comes from the fact that the exam is considered passed from 35\%, and larger results are offset by the formula linearly. Nevertheless, BingAI had problems with several questions in this exam and either answered incorrectly or omitted information. StableLM-7B and LLaMa-7B-Q had no chance of passing this exam, with a performance of 6.7\% and 13.3\%, respectively. Nearly every answer had massive errors or large information gaps. The models also lost context in between and started talking about completely different topics.

In the simulated oral exam on numerical analysis, mainly comprehension questions were asked, and hardly any calculations had to be done. This led to excellent results for both GPT models. ChatGPT-3.5 got a grade of 1.6, with 90\% of the total score, whereas GPT-4.0 increased this to 95\% and a grade of 1.3. Both models could answer almost every question completely and correctly and only made minimal errors. BingAI had great problems in this exam, although it was mainly about knowledge reproduction, and scored well behind the GPT models with 68\% of the total score, or a grade of 2.9. BingAI had problems with several questions and made mistakes while reproducing information. For example, when asked about the complexity of the \textit{Gauss Algorithm}, BingAI gave a reference to Wikipedia but then misquoted the article with a complexity of O(n²).

\subsection{5th Semester}

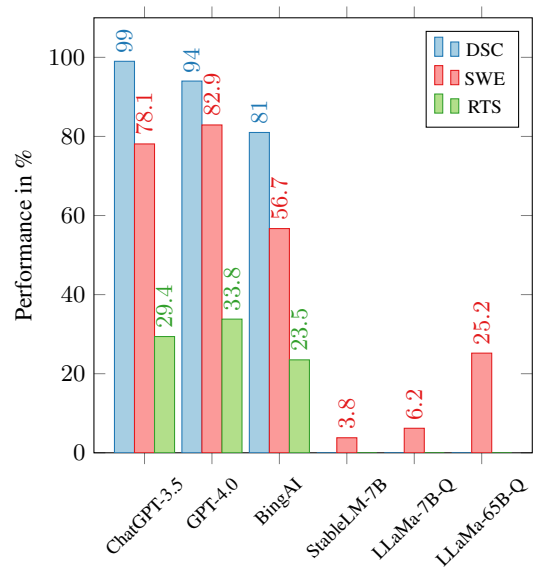
\begin{figure}[t]
	\centering
	\begin{tikzpicture}
		\begin{axis}[
			ybar=2*\pgflinewidth,
			symbolic x coords={ChatGPT-3.5, GPT-4.0, BingAI, StableLM-7B, LLaMa-7B-Q, LLaMa-65B-Q},
			enlarge x limits=0.15,
			ybar=0pt,
			xtick=data,
			ylabel=Performance in \%,
			nodes near coords,
			nodes near coords style={rotate=90, anchor=west},
			ymin=0,
			ymax=110,
			bar width=7.5pt,
			cycle list name = mgrefColorFilled,
			width=0.85\linewidth,
			height=0.85\linewidth,
			legend style={font=\scriptsize},
			legend pos = north east,
			xticklabel style={font=\scriptsize, rotate=45, align=center},
			xlabel style={align=center},
			nodes near coords={%
				\pgfmathfloatifflags{\pgfplotspointmeta}{0}{}{\pgfmathprintnumber{\pgfplotspointmeta}}},
			]
			\addplot coordinates {(ChatGPT-3.5, 99) 	(GPT-4.0, 94) 	(BingAI, 81) 	(StableLM-7B, 0) 	(LLaMa-7B-Q, 0) 	(LLaMa-65B-Q, 0)};
			\addplot coordinates {(ChatGPT-3.5, 78.1) 	(GPT-4.0, 82.9) (BingAI, 56.7) 	(StableLM-7B, 3.8) 	(LLaMa-7B-Q, 6.2) 	(LLaMa-65B-Q, 25.2)};
			\addplot coordinates {(ChatGPT-3.5, 29.4) 	(GPT-4.0, 33.8) (BingAI, 23.5) 	(StableLM-7B, 0) 	(LLaMa-7B-Q, 0) 	(LLaMa-65B-Q, 0)};
			\legend{DSC, SWE, RTS}
		\end{axis}
	\end{tikzpicture}
	\caption{Exam results for Data Science (DSC), Software Engineering (SWE) and Real-Time Systems (RTS)}
	\label{fig:Semester_5_Performance}
\end{figure}

We have received approval for three modules of the fifth semester: Data Science, Software Engineering, and Real-Time Systems. The results can be seen in Figure \ref{fig:Semester_5_Performance}.

Data Science (DSC) is an elective course with five credit points. The module provides an introduction to Big Data and Machine Learning. Students of this module will learn to extract, prepare and analyze large data sets \cite{modulhandbuch_po2013}. The module concludes with an oral exam which we simulated by taking questions from a questionnaire.

Software Engineering (SWE) is a five credit-point module. It covers advanced solutions for building, testing, and maintaining large IT systems and techniques for organizing big software projects. A special focus is on effort estimation and (agile) software development processes \cite{modulhandbuch_po2013}. As determined by the supervising professor, the module is considered to be passed if 33\% of the total points are achieved.

Real-Time Systems (RTS) is a five credit-point module. It focuses on the architecture, the concepts, and the functionalities of modern real-time systems. Students learn aspects of concurrent real-time programming and how to deal with time constraints and task management \cite{modulhandbuch_po2013}. The module is concluded with a written exam in which a special focus is placed on manual real-time proof for various scheduling methods.\\

DSC is the best-performing module for ChatGPT-3.5 in all our tests. With 99\%, ChatGPT-3.5 scored a 1.0. Every question was answered correctly, from problems in the field of \textit{Big Data} to data preparation, to classification and clustering methods. DSC is also the only module in which ChatGPT-3.5 outperformed GPT-4.0. The latter model did not score itself badly with 94\% and a score of 1.3, but unfortunately gave partially wrong answers to questions about \textit{Eventual Consistency} and \textit{Sharding}. Such results are possible since we only prompt all models once in our tests. BingAI performed again worse than the GPT models. With 81\%, it achieved a score of 2.1. It was challenging for BingAI to make its own decisions in tasks, e.g., choosing between an aggregate-oriented or a relational data model.

In SWE, the GPT models were again well ahead of BingAI. ChatGPT-3.5 achieved a 1.9 with 78.1\%, GPT-4.0 a 1.7 with 82.9\%, while BingAI only achieved a 2.9 with 56.7\%. The GPT models answered most of the questions completely and correctly but made massive errors in designing test cases for a finite state machine. BingAI could not solve this task either and made errors in explaining design patterns and performing an effort estimation using \textit{Function Point Analysis}. StableLM-7B and LLaMa-7B-Q performed poorly in this module, scoring 3.8\% and 6.2\%, respectively. LLaMa-65B-Q performed significantly better than the 7 billion parameter version and nearly passed the exam with 25.2\%, out of 33\% needed. Interestingly, LLaMa-65B-Q was able to partially answer difficult questions on software development principles and the design of component diagrams but failed to explain unit tests.

RTS is a demanding exam in which many calculations have to be done. Computational time requirements, core workloads, and a large, manual real-time proof must be calculated. This exam has proven to be extremely difficult for all models tested. ChatGPT-3.5 scored 29.4\%, GPT-4.0 scored 33.8\%, and BingAI scored 23.5\%. Accordingly, all models failed the exam. The models could answer a few simple introductory questions but, early on, miscalculated the computational kernel allocation for a round-robin scheduling procedure. No model was able to solve this task correctly. Likewise, no model was able to calculate the real-time proof correctly. This task is nested, with each intermediate calculation evaluated individually but often a prerequisite for the next calculation. BingAI lost all context in this task after the third partial calculation. The GPT models could continue to calculate but miscalculated fatally early on. Both the calculation path and the result were not correct. When calculating the average execution time for processes of a machine, both ChatGPT-3.5 and GPT-4.0 set up the correct formula, adding all times and dividing by the amount, but only GPT-4.0 also got the correct result; ChatGPT-3.5 miscalculated. 

All in all, the models were heavily overcharged with this exam.

\section{Discussion}

An overview of the average performances and results achieved across all tested modules can be seen in Table \ref{table:Performance_summary}. It is important to note that all data collected is only a snapshot that considers the systems' performance at the time of the assessment. These systems continue to evolve.

ChatGPT-3.5 achieved an average of 79.9\% of the maximum possible score in the ten modules tested. ChatGPT-3.5 performed particularly well in modules with a high proportion of web development or high-level programming language content, like Python and JavaScript. Even in the field of data science, ChatGPT-3.5 achieved almost full marks. In exams with more complex tasks like Operating Systems or Data Network Management, ChatGPT-3.5 often provided at least an approach to the solution. We noted major difficulties for this LLM with various tasks that required mathematical calculations. The application of scheduling algorithms and the calculation of core utilization and process runtimes posed significant challenges for ChatGPT-3.5. Due to these shortcomings, passing the Real-Time Systems exam is not currently possible. Accordingly, the LLM would not be capable of completely finishing our bachelor's degree program in computer science. However, with an understanding of its strengths and weaknesses, ChatGPT-3.5 shows great potential as an outstanding learning aid for students and lecturers.

GPT-4.0 achieved even better results than ChatGPT-3.5, obtaining an average performance of 80.2\%. This score is expected to increase even further if the missing modules are tested with GPT-4.0. A strong focus on specific programming languages or fields of computer science, as observed in ChatGPT-3.5, could not be detected in GPT-4.0. At the same time, GPT-4.0 demonstrated a more consistent overall performance. Like ChatGPT-3.5, GPT-4.0 had difficulties with tasks that required calculations; this also resulted in the failure of the Real-Time Systems exams. For the same reason, GPT-4.0 would not be able to finish the degree program. However, it should be noted that GPT-4.0, despite the identified difficulties, represents an improvement in all areas over ChatGPT-3.5. The use of plugins, for instance, to redirect mathematical computations to a system like WolframAlpha could significantly improve this outcome. This is left to be explored in future studies.

BingAI scored much lower than the GPT models in our tests, with 68.4\%. It was the only one of these three models that failed two exams rather than just one, with one of the exams (Data Network Management) being not calculation-intensive. BingAI often encountered problems when the solutions were not directly searchable online. Thus, it made mistakes in extracting information from texts or creating and presenting solutions. Even when the answer to a question could be found via an internet search, BingAI sometimes made inexplicable citation errors. BingAI also provided the shortest responses of all the LLM systems tested, often ignoring aspects of a question. According to the current state, BingAI is inferior to the GPT models.

LLaMa-7B-Q showed poor results, with an average performance of 12.3\% in six tested modules. It often had difficulties understanding questions, lost context, or started talking about completely different topics. LLaMa-7B-Q could not solve a single task of an exam. According to our tests, it would not be possible for this LLM to pass any module.

LLaMa-65B-Q showed better results with an average performance of 20.0\%, but was also tested only in two modules. It scored one percentage point and 19 percentage points better than its 7 billion parameter counterpart. At this point, more tests are needed to make a final statement about the performance differences between these models. Nevertheless, a trend can be determined: LLaMa-65B-Q performs significantly worse than BingAI, let alone the GPT models. After our tests, whether it would pass a single module is questionable, and it is not suitable for use as a learning aid.

\begin{table}[t]
	\begin{center}
		{\caption{Summary of the average performance\\ proportionally calculated to all modules taken.}\label{table:Performance_summary}}
		\begin{tabular}{cccc}
			\hline
			\\[-6pt]
			\textbf{Model} & \textbf{Average score} & \textbf{\# Modules} & \textbf{Passed / Failed}\\
			\hline
			\\[-6pt]
			\quad GPT-4.0 		& 80.2\% & 6  & 5 / 1\\
			\quad ChatGPT-3.5 	& 79.9\% & 10 & 9 / 1\\
			\quad BingAI 		& 68.4\% & 10 & 8 / 2\\
			\quad LLaMa-65B-Q 	& 20.0\% & 2  & 0 / 2\\
			\quad LLaMa-7B-Q 	& 12.3\% & 6  & 0 / 6\\
			\quad StableLM-7B 	& 10.8\% & 6  & 0 / 6\\
			\hline
		\end{tabular}
		\vspace{-12pt}
	\end{center}
\end{table}

StableLM-7B, tested in six modules, achieved the worst results of all tested LLMs with 10.8\%. It was unable to answer any question correctly and completely. Interestingly, StableLM-7B often related questions to a business context or attempted to answer them in such a context. StableLM-7B even understood complex questions from the field of project management but could not establish a reference to computer science or software development. According to our tests, this LLM is also unsuitable as a learning aid.

\section{Conclusion}
In the presented study, we tested and evaluated the performance of various LLMs across a series of modules in a bachelor's computer science degree program. Our results are in line with existing research (e.g. \cite{binz2023using}) by showing strong performances of Generative Pre-training models (GPT) across an undergraduate curriculum while having severe restrictions in key areas.

A prevalent worry is the potential for essays to progressively lose significance as evaluation tools within higher education \cite{Gimpel2023Unlocking}. Our tests show the strength and topic affinities of current LLMs, but also their weaknesses, distinctively in mathematical computations. We conclude from these results that a comprehensive blueprint for our curriculum remains elusive at this point. Despite this, the deployment of these models presents lecturers with challenges, as the detection of plagiarism in AI-generated content is not particularly mature yet \cite{khalil2023chatgpt}. It is imperative to recognize that the GPT models in our tests have completed numerous examinations with scores above 95\%. Given that some of our examination rules allow aids, and the pattern of past exams often remains unchanged, the sophisticated capabilities of these models could potentially create near-perfect and legally permissible "cheat" sheets. This, combined with the advancing abilities of current LLMs \cite{Gimpel2023Unlocking}, compels us to reconsider and construct robust examination methods. Oral and written exams without aids remain valid alternative options \cite{Gimpel2023Unlocking}.

The smaller models in our tests exhibit substantial performance deficiencies, with profound disparities encountered in almost all performance-defining areas. Consequently, they currently do not measure up as viable educational aids.

Continued research may examine the performance of existing models in unexplored curriculum modules. Furthermore, additional modules could be examined to provide a broader overview. Future research could also extend to the study of other LLMs, such as \href{https://bard.google.com/}{Google Bard}. Also, broadening the scope to related disciplines, like electrical engineering, would be beneficial to gain a better understanding of domain-specific performance capabilities.

\ack{The authors declare that funds of the \textit{Bundesministerium für Bildung und Forschung} were used to finance this study.\\ Grand-ID: 16DHBKI070}

\bibliography{ecai}
\end{document}